# Robust edge photocurrent response on layered Type II Weyl semimetal WTe$_2$

Authors: Qinsheng Wang[1,2], Jingchuan Zheng[1,2], Yuan He[1,2], Jin Cao[1], Xin Liu[3], Maoyuan Wang[1], Junchao Ma[3], Jiawei Lai[3], Hong Lu[3], Shuang Jia[3,4], Dayu Yan[5], Y.-G. Shi[5], Junxi Duan[1,2], Junfeng Han[1,2], Wende Xiao[1,2], Jian-Hao Chen[3,4], Kai Sun[6], Yugui Yao[1,2]*, Dong Sun[3,4]*

Affiliations:

[1]Key laboratory of advanced optoelectronic quantum architecture and measurement (MOE), School of Physics, Beijing Institute of Technology, Beijing, China

[2]Micronano Centre, Beijing Key Lab of Nanophotonics & Ultrafine Optoelectronic Systems, Beijing Institute of Technology, Beijing, China

[3]International Center for Quantum Materials, School of Physics, Peking University, Beijing, China.

[4]Collaborative Innovation Center of Quantum Matter, Beijing, China.

[5]Beijing National Laboratory for Condensed Matter Physics, Institute of Physics, Chinese Academy of Sciences, Beijing, China

[6]Department of Physics, University of Michigan, Ann Arbor, Michigan 48109-1040, USA

Correspondence to: ygyao@bit.edu.cn; sundong@pku.edu.cn

**Abstract:**

Photo sensing and energy harvesting based on exotic properties of quantum materials and new operation principles have great potentials to break the fundamental performance limit of conventional photodetectors and solar cells. As topological nontrivial materials, Weyl semimetals have demonstrated novel optoelectronic properties that promise potential applications in photo detection and energy harvesting arising from their gapless linear dispersion near Weyl nodes[1-5] and Berry field enhanced nonlinear optical effect at the vicinity of Weyl nodes[6-9]. In this work, we demonstrate robust photocurrent generation from charge separation of photoexctied electron-hole pairs at the edge of $T_d$-WTe$_2$, a type-II Weyl semimetal, due to crystalline-symmetry breaking along certain crystal fracture directions and possibly enhanced by robust fermi-arc type surface states[1,3,10]. Using scanning photocurrent microscopy (SPCM) measurements, we further demonstrate that the edge current response is robust over a wide excitation photon energy. We find that this robust feature is highly generic, and shall arise universally in a wide class of quantum materials with similar crystal symmetries. In addition, possible connections between these edge photocurrents and topological properties of Weyl semimetals are explored. The robust and generic edge current response demonstrated in this work provides a new type of charge separation mechanism for photosensing and energy harvesting over broad wavelength range.

**Introduction**

Efficient separation of photo-excited electron-hole pairs is an essential ingredient for various highly applicable fields such as photo sensing[11], solar energy harvesting[12] and photo catalysis[13]. For this purpose, imbalanced momentum distribution of non-equilibrium photo-excited charge carriers, either from population imbalance or velocity imbalance in k-space, is required. Technically, this can be achieved by breaking inversion symmetry either internally or externally in devices. In conventional semiconductors with inversion symmetry, such as silicon, GaAs and graphene, applying a bias voltage over the electrodes externally is a straightforward approach to break the inversion symmetry of the device. Advanced approaches of inversion symmetry breaking through external ways can also be achieved by introducing a built-in electric field in a PN junction[14], Schottky junction[15] or at interface with other materials, or domains processing different Seebeck coefficients through photo thermoelectric effect (PTE)[16]. Although the above schemes have been successfully demonstrated for photo-excited charge separation, there are fundamental limitations on device fabrication and specific applicable circumstances. For example, applying an external bias on detectors based on semi-metallic materials such as Dirac/Weyl semimetals is not feasible because the external bias can induce a significant current, termed as the dark current, in conducting materials even in the absence of light[11,17,18]. PN junctions can efficiently separate photoexcited electron-hole pairs, but its application is limited by the requirement of doping, which can be very challenging for certain materials.

On the other hand, the inversion symmetry breaking can naturally occur in some classes of crystals, which could lead to efficient charge separation of photo-excited carriers and result in shift current response[19]. This phenomenon, which is often referred to as bulk photovoltaic effect (BPVE) or anomalous photovoltaic effect[20,21], is due to the fact that the real space position of electron shifts a little as the expected electron positions of the ground state and the excited state are slightly different after photoexcitation under homogenous photo illumination in non-centrosymmetric crystal. Previously, this effect is mainly demonstrated in semiconductors with non-centrosymmetric structure such as $LiNbO_3$[22] and $BaTiO_3$[23]. Nevertheless, most of these crystals have large bandgaps, which limit their applications in low-energy photo detection and energy harvesting. Weyl semimetals with inversion symmetry breaking provide a promising class of material candidates via this route. The gapless linear-dispersed band structure of Weyl semimetal allows broadband wavelength operation extendable to far-infrared or even Terahertz range. In addition, the monopole-type of Berry curvature around the Weyl nodes can enhance the nonlinear optical effects at the vicinity of Weyl nodes[7], which leads to a large photocurrent response over a broad wavelength range[24]. However, the shift current response in inversion-symmetry-breaking Weyl semimetals is cancelled under the light incident geometry that the surface perpendicular to the incident light still preserves the in-plane inversion symmetry. For example, in well-established Type-II Weyl semimetals with $C_{2v}$ crystal point group, such as $WTe_2$, $MoTe_2$ and $TaIrTe_4$[25,26], the second order nonlinear shift current response vanishes if light incidence direction is along the crystallographic c-axis[25]. In this work, we experimentally show that the photo excited electron-hole pairs can be separated efficiently along certain edges due to the broken of $C_{2v}$ symmetry upon fracturing along certain crystallographic directions, which provides robust edge current response in Type-II Weyl semimetal $WTe_2$ under normal incidence. Interestingly, although the survival of the photocurrent response is solely determined by the edge fracture direction from a pure crystal symmetry consideration, the observed responsivity possibly relates to the fermi-arc type surface states.

$T_d$-$WTe_2$ with crystal structure shown in Figure 1a is proposed as an inversion-symmetry breaking type-II Weyl semimetal[27]. It belongs to space group of $Pmn2_1$(No. 31) and point group $C_{2v}$, which consists a two-fold rotation symmetry with screw rotation axis $C_2$ along the crystallographic $\hat{c}$-axis, and mirror symmetry with mirror plane ($M_a$) perpendicular to the crystallographic $\hat{a}$-axis and glide mirror symmetry with mirror plane ($M_b$) perpendicular to the crystallographic $\hat{b}$-axis. In the first Brillion zone, $WTe_2$ has 8 Weyl nodes laying on $K_c$=0 surface, as marked in Figure 3d. These Weyl nodes come in two quartets located 0.052 eV and 0.058 eV above the Fermi level, all of which are type-II Weyl nodes that exist at the boundaries between electron and hole pockets[27]. Due to the inversion symmetry breaking, $WTe_2$ demonstrates large second order photocurrent response in Weyl semimetals with oblique light excitation[25], but for normal incident light excitation with electric fields lying in the $\hat{a} - \hat{b}$ plane, the planar components of a rank-3 tensor with the $C_{2v}$ symmetries are all null: $T_{\alpha\beta\gamma} = 0$, due to the existence of in-plane inversion symmetry in the $\hat{a} - \hat{b}$ plane.

In a typical scanning photocurrent measurement shown in Figure 1b, 633-nm pulse filtered from a fiber white light super-continuum source is used to excite a 6-nm thick two-terminal $WTe_2$ device under un-biased condition at room temperature. According to the scanning photocurrent response shown in Figure 1c, d, the photocurrent is mainly generated either along the interface of $WTe_2$-metal contact or along the edge of the $WTe_2$ flakes. The response at the $WTe_2$-contact interface is expected, which has been well understood from previously studies on similar devices based on other 2D layered semimetals[26,28]. The charge separation is induced by the interplay of built-in electric field and PTE effect at the interface in absence of external voltage bias. When the excitation is away from the contact area, the major charge separation mechanisms (eg. PTE, built-in electric field, photo Dember) all vanish, thus the PC response should be negligible because all the planar components of rank-3 tensor are zero under normal incident light excitation. This is consistent with the experimental observation that no photocurrent is observed over the region away from the metal contact (Figure 1d). However, we have observed clear photocurrent response of 27 μA/W at the edge of the device which is totally unexpected. Such robust edge photocurrent response, to the best of our knowledge, has never been observed without applying an external magnetic field in non-chiral materials[29,30].

To further investigate the edge current response of $T_d$-$WTe_2$, 8 devices are fabricated with edges fracturing along different crystal orientations within a-b plane. The SPCM results of three representative devices are shown in Figure 2. The in-plane fracture is random during the mechanical exfoliation process, but for $WTe_2$, it has the tendency to fracture along certain in-plane crystallography orientations[31]. The most preferable fracture direction within a-b plane is along the tungsten chain direction (<100>). Experimentally, the crystal orientation of the edge can be identified by polarization resolved Raman spectroscopy, the ratio of Raman peaks at 164 and 212 cm$^{-1}$ reaches maximum when the laser polarization is parallel to the <100> direction as shown in Section S3 of supplementary information[32].

An obvious feature from Figure 2 is that not all edges have photocurrent responses, photocurrent generation only happens on edges along certain crystal orientations. Even within one device, some edges exhibit photo responses while the other edges don't. This is demonstrated in the 150-nm thick device shown in Figure 2a. The device has irregular top edges with 5 different major directions and the long bottom edge is identified to be along the <100> direction. According to the SPCM shown in Figure 2b, we observe photocurrents along 3 of the 5 edges at the top, while the other 2 top edges and the bottom edge show no photocurrent responses, and all these edges with vanishing photocurrent response are identified to be along the <100> direction. Furthermore, the photocurrent from the area labelled 1 on the top edge is negative, while those from the areas labelled 3 and 5 are positive. The device shown in Figure 2c and 2d exhibits similar but more salient photocurrent response. The top edge along <100> has no photocurrent response, while the two bottom edges, with a fracture direction about 11-degree angle with respect to the top edge, show clear photocurrent responses.

The third device shown in Figure 2e-h provides an additional feature of the edge current response: the edge photocurrent response may switch sign over different edges of one device. The SPCM between contacts 1 and 2, contacts 2 and 3, contacts 4 and 5 are shown in Figure 2 f, g, h respectively (more results from other contacts shown in supplementary information S4). The long edge on the left (Figure 2e), which are along the <100> direction, has no photocurrent response. The right edge in Figure 2g, which is parallel to the long left edge, doesn't show any edge current response either. The edges along lower symmetry directions in Figure 2f have photocurrent response, the response of different edges can have opposite photocurrent sign. In Figure 2h, besides the free-edge current responses, the photocurrent can also be generated on multilayer steps as marked by the dash circles (similar photo response from step-edge marked by dash circle is also observed in Figure 2f). AFM measurement indicates the step is about 80 nm (supplementary information S2). From the above experimental results, we conclude that the edge photocurrent response is determined by the crystal fracture orientation of the edge: regardless of the flake thickness, there is no edge photocurrent response if the edge fractures along the high symmetry <100> or <010> direction (supplementary information S9). However, for low symmetry fracture orientations that are not parallel to the <100> or <010> direction, there are clear edge current responses.

The crystal symmetry of the step-edges is lowered in a very similar way to free edges, and thus the photocurrent response from the step-edge shares the same origin as the photocurrent response of free-edge.

The emergence of the edge current response along the low symmetry fracture edges observed in Figure 2 can be well explained through the following symmetry consideration. The process of photocurrent generation can be described as:

$$j^\alpha = \sigma^{\alpha\beta\gamma} E^\beta(\omega) E^\gamma(-\omega) \tag{1}$$

where $j$ is the DC current density and $E(\omega)$ is the optical electric field, where indices $\alpha$, $\beta$, and $\gamma$ mark spatial components of these vectors. The coefficient $\sigma^{\alpha\beta\gamma}$ must obey the symmetry of the crystal, and it shall vanish unless the two sides of the equation have identical response to all symmetry operations, known as the symmetry selection rule, which can be systematically constructed utilizing group representations[33]. For a crystal with inversion symmetry, because the left hand side (l.h.s.) $j^\alpha$ is odd under space inversion, while $E^\beta(\omega) E^\gamma(-\omega)$ on the right hand side (r.h.s.) is even, the coefficient $\sigma^{\alpha\beta\gamma}$ must vanish, i.e., this photocurrent is prohibited by symmetry. For a 2D system, the photocurrent also vanishes if there is a 2-fold axis perpendicular to the 2D plane. This is because, as far as the in-plane components of $j$ and $E(\omega)$ are concentered, under such a 2-fold rotation, both these two quantities change sign, and thus again the l.h.s. and r.h.s. of the equation get opposite signs, making $\sigma^{\alpha\beta\gamma} = 0$. This is the key reason why photocurrent is absent in the bulk in our experiment, despite that the 3D space-inversion symmetry is already explicitly broken by the crystal structure. Along the edge, however, the 2-fold rotational symmetry is broken, and thus the selection rule that prohibits the photocurrent is now lifted, which results in the emergence of edge photocurrents that we observed. Although the break of 2-fold rotational symmetry at the edge in general allows a photocurrent, other symmetries may enforce further constraints on the direction of the current. In particular, if the edge is perpendicular to a mirror plane of the crystal, eg. <100> and <010>, this mirror symmetry will be preserved by the edge. Because the r.h.s. of the equation is invariant under mirror, only current parallel to the mirror (i.e. invariant under the mirror) is allowed by symmetry. In our experiment, this means that if we consider edges along the <100> direction (Figure 3a), the mirror plane perpendicular to the edge, along <010>, will confine the photocurrent along the <010> direction, while the current response along the <100> direction stays zero. According to the Shockley-Ramo theorem illustrated below, which describes the nonlocal photocurrent response from local charge separation, the local photocurrent along the perpendicular direction to the edge cannot induce a macroscopic signal that is perpendicular to the weighting field. The Shockley-Ramo theorem is based on the fact that the current induced in the electrode is due to the instantaneous change of electrostatic flux lines into or out of the electrode. The macroscopic measured current I at the electrode can be described by the local excited current density $j_{loc}(x,y)$ as[29]:

$$I = A \int j_{loc}(x,y) \cdot \nabla \emptyset(x,y) dx dy \qquad (2)$$

where the $\nabla \emptyset(x,y)$ is an auxiliary weighting field, which is the solution to a suitable Laplace problem assuming that $\emptyset = 0$ at the ground contact and $\emptyset = 1$ at the current-collecting contact; A is a prefactor that depends on the device configuration. We provide the simulation results of the weighting field lines near the edge of the device as shown in Figure 3c, which is nearly parallel to the edge. So for $j_{loc}$ perpendicular to the edge, the response is zero. This is consistent with our experimental observation that the edge current response vanishes when the edge fractures along the <100> direction. For edges oblique to the mirror planes (Figure 3b), because the mirror symmetry is broken along these edges, $j_{loc}$ along and perpendicular to the edge are both allowed and thus a macroscopic photocurrent signal can be observed. We note our discussion applies to relatively thick samples. For thinner layers when the translational symmetry along c is effectively broken, the glide mirror plane $M_b$ and the rotational symmetry around c should disappear, which may result in different edge response that requires further investigations[34,35].

To further elucidate the edge current generation mechanism, we have performed systematic excitation power, polarization and wavelength dependent studies. Figure 4a shows the magnitude of edge current response is linear with the excitation power, which indicates the edge photocurrent generation is a second order process of excitation light field (I~E(-ω)E(ω)). The response has no clear differences in terms of SPCM response pattern for horizontal and vertical polarized light excitation as shown in Figure 4b & 4c. Further linear polarization dependent measurement indicates the response is slightly anisotropic (Figure 4d): the long axis of the anisotropic ellipse is along the <100> direction, and the anisotropy ratio is 1.1. The insensitive photocurrent response on the excitation light polarization is consistent with the shift current nature[36] of the photocurrent generated on the edge. In Figure 4e-h, we note that the SPCM pattern is quite inert to the excitation wavelength over a broad photon energy range from 0.12 - 1.96 eV. The response pattern is qualitatively the same on the edge except that the spatial resolution degrades at longer wavelength due to wavelength dependent diffraction limit. The same direction of the generated photocurrent over difference wavelengths indicates the sign of the second order nonlinear DC photoconductivity tensor $\sigma^{\alpha\beta\gamma}(0; \omega, -\omega)$ stays the same for different wavelengths.

It is worthwhile to emphasize that the edge current generation is in contrast to the conventional approach, where an external voltage is needed to separate electron-hole pairs and to obtain a current. In the edge current response, the crystal itself provides an effective electric field near the edge of the sample originating from its own symmetry, which naturally separates electron-hole pairs without external input or the needs to construct sophisticated heterostructures. This symmetry analysis, as well as the conclusions, applies generically to any materials with similar symmetries, independent of any microscopic details or the topological properties of the materials. Although the symmetry fully dictates the qualitative behavior of the observed phenomena and guarantees the existence of the edge photocurrent, it is worthwhile to emphasize that symmetry alone doesn't predict the magnitude of the photo response, which relies on material properties. In particular, because the significant edge photocurrent is observed in a topologically nontrivial Weyl semimetal, which has nontrivial topological features at the edge/surface, it is not unreasonable to suspect whether the topology surface states are responsible for the significant response amplitude. To have a conclusive answer, more detailed investigations are needed, which is beyond the central focus of our study. However, theoretically, there exists certain mechanisms, which could result in constructive interplay between the topological edge modes and the edge photocurrent, and thus enhance the amplitude of the photocurrent. For example, the topological surface states offer one additional conducting channel, on top of the bulk contributions. Interestingly, the density of states of this extra (surface/edge) conducting channel depends on the edge orientation in the same way as the edge photocurrent, i.e. it vanishes on the (100) and (010) surfaces where the edge photocurrent is zero, but remains finite on surfaces with high Miller indices, where the photocurrent is observed. This is illustrated in more detail in Figure 3. In consistent with the symmetry consideration shown in Figure 3a,b, for high symmetry facture edge along the <100> or <010> direction, the projection of Weyl nodes with opposite chirality annihilate in pairs as projected onto the (100) and (010) surface (Figure 3d). Because a Fermi arc of Weyl semimetal has to start from the projection of Weyl node of one chirality on the surface and terminate on the projection of Weyl node of the opposite chirality, there is no Fermi arc along these edges. However, the Fermi arc can possibly survive in higher Miller-index surface Brillouin zones (BZ), for example, when projected onto the (110) surface. In addition, as shown above, one key ingredient of edge photocurrents is the symmetry breaking effect induced by the 1D edge. Although geometrically the edge is a 1D line, its symmetry breaking effect is not limited to the first line of atoms at the edge. Instead, it shall penetrate into the bulk by some distance. Longer penetration here means that a wider range near the edge will be influenced by edge and shows signature of broken 2-fold symmetry, and thus results in a stronger edge photocurrent. For Weyl semimetal, topology offers an extra channel for the information about the edge to penetrate into the bulk, through the topologically-protected edge states. The quantum wavefunction of an edge state penetrates into the bulk, and its penetration depth varies according to the surface momentum[37,38], which helps spreading information about the edge (e.g. symmetry reduction at the edge) to a wider region into the bulk, and thus enhance the intensity of the edge photocurrents.

In summary, we have observed a robust photocurrent response on the edge of $T_d$-$WTe_2$ along low symmetry crystal fracture directions due to the lifting of symmetry constraints. The Fermi-arc type surface states may play a role in the edge photocurrent response although the contribution of these states requires further investigations. The work resolves the photocurrent response under different symmetry conditions of the edge and imply the important roles of the edge or the surface in the photo response of topological semimetals. We note the edge photocurrent response provides a new type of charge separation mechanism for photosensing over broad wavelength range and it is quite universal and applicable to a wide class of quantum materials with similar symmetries. This finding enables charge separation through the engineering of edge and surface in quantum materials, especially that in topological semimetals.

**Methods:**

**Device fabrication:**

The $WTe_2$ flakes were mechanical exfoliated onto a silicon substrate with a 285-nm oxide layer. Edges of different fracture directions were formed randomly during the mechanical exfoliation process. The electrode was patterned by e-beam lithography and then 5/100 nm Cr/Au were evaporated for electric contacts. For devices with relatively thin thickness, a PMMA layer was spin-coated on the device to protect the device from degradation under ambient conditions. For thick devices (>30 nm), the measurements were performed under ambient conditions.

**Optoelectronic Measurements:**

Standard scanning photocurrent measurements[16] were performed in ambient conditions at room temperature using a 633 nm pulse laser with a ~0.85 μm spatial resolution. The reflection signal and photocurrent were recorded simultaneously to get the reflection and photocurrent mapping. The laser beam was modulated with a mechanical chopper (2833 Hz), and the short-circuit photocurrent signal was detected with a current preamplifier and a lock-in amplifier. The sample was mounted on a three dimensional piezo stage for scanning measurement. For wavelength-dependent measurements, an acousto-optic tunable filter was used to select the desired wavelength from white light supercontinuum output of YSL Photonics laser (Model number SC-Pro, 4.5 MHz, 100 ps, 450 − 2400 nm). The laser was focused on the same spot of the device by a 100× NIR objective. For 4 μm and 10.6 μm response measurement, two continuum wave quantum cascade lasers(QCL) were used. The lasers were focused on the sample by a 40× reflective objective (NA = 0.5), and the light spot size was estimated to be 10 μm for 4-μm QCL and 20 μm for 10.6-μm QCL, evaluated from a scanning reflection image on the sample and electrode.

**Data availability:**

All data supporting the findings in this manuscript are available from the corresponding authors (Y.Y. and D.S.) upon request.

**Figures and Captions**

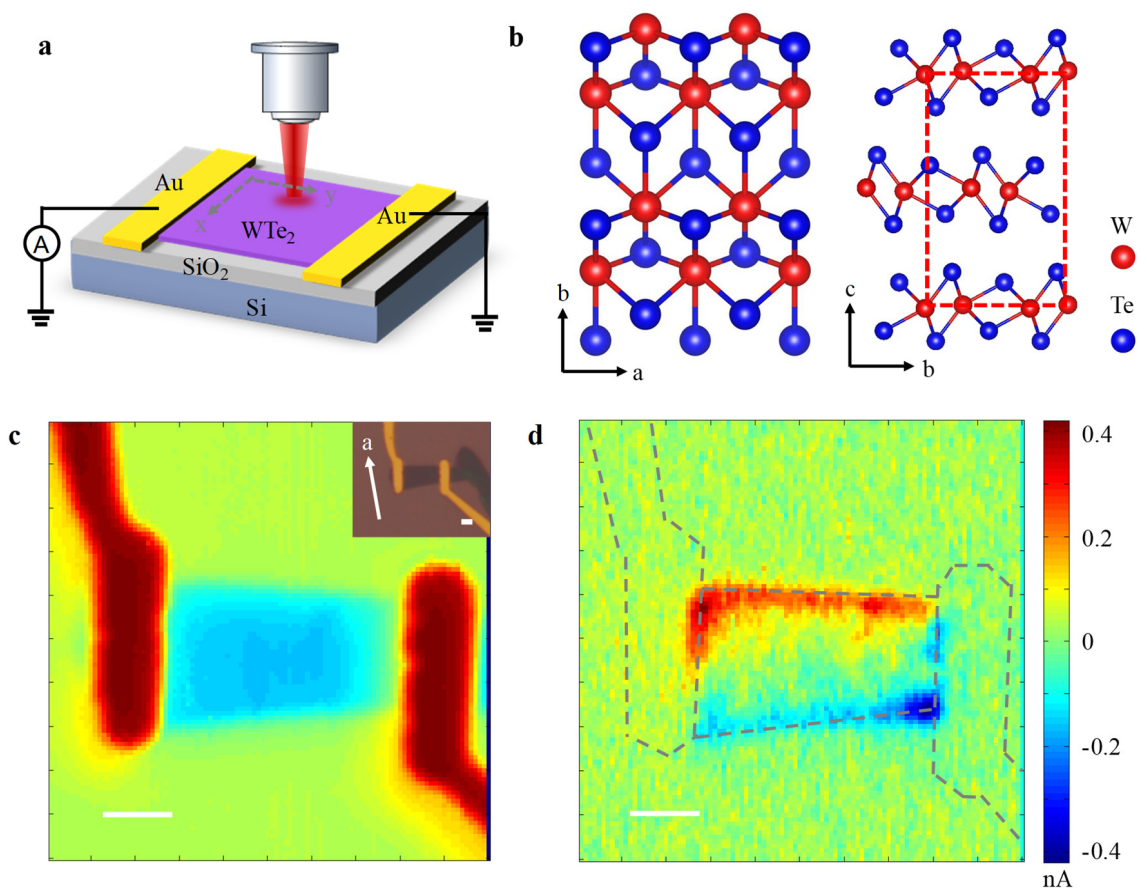

**Figure 1| Edge photocurrent response in WTe₂. a**, Schematic diagram of scanning photocurrent measurement of a WTe₂ field effect device. **b**, Crystal structure of WTe₂. Top view of the crystal structure of a-b plane (left) and cross-section view of the interlayer stacking structure (right) of T$_d$- WTe₂. **c, d,** Scanning reflection image (c) and scanning photocurrent response (d) of a typical 6-nm thick WTe₂ device. The measurement was performed with 633-nm light filtered from white light super-continuum and the power shine on the sample was about 15 μW. The excitation light was focused using a 100X objective lens and the focused spot size was 0.85 μm. The inset of Figure c is the optical microscope image and the white arrow marks the crystallographic a-axis. The scale bars are all 2 μm.

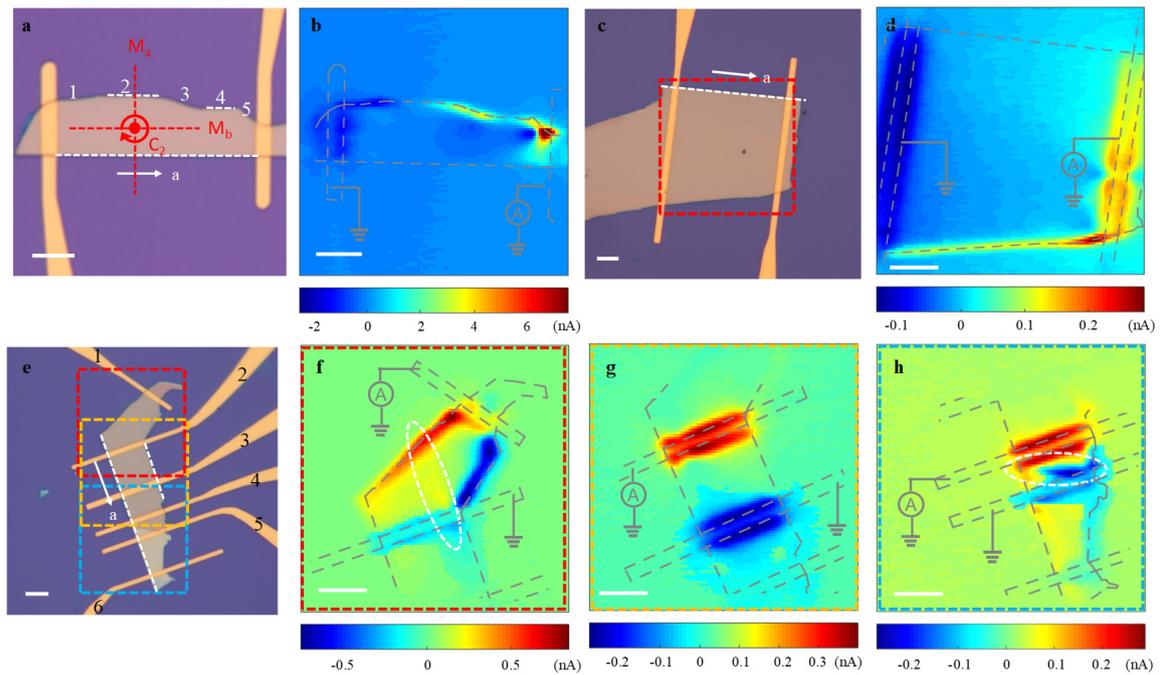

**Figure 2| Photocurrent response from edges along different crystal fracture directions. a** and **b**, Optical microscope image (a) and scanning photocurrent image (b) of a device with irregular shape. **c** and **d**, Optical microscope image (c) and scanning photocurrent image (d) of a device with trapezium shape. **e, f, g** and **h**, Optical microscope image (e) and scanning photocurrent image of different parts (f, g and h) of a long device with edges along various crystal fracture directions. All measurements were excited with 180-μW 1.96-eV pulse laser. The white arrows in microscopy images mark the crystallographic a-axis and the white dashed lines mark the edges that have no photocurrent responses. The red symbols in Figure (a) display the point group symmetries of the crystal lattice. Photocurrents only arise along edges that break the mirror symmetry, in good agreement with the symmetry analysis. The scale bars are all 8 μm.

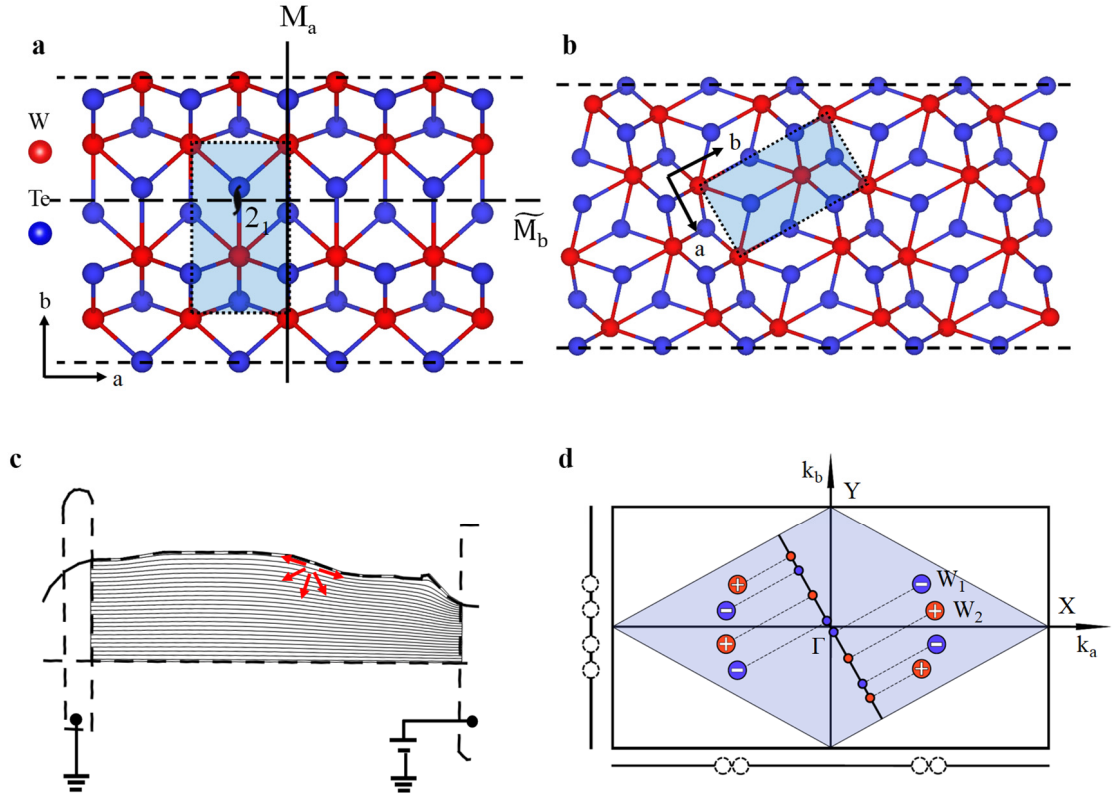

**Figure 3| Symmetry analysis of edges along different crystal fracture directions of WTe$_2$. a and b,** Schematic of lattice structures with edges along <100> (a) and <110> (b) directions respectively. The unit cell is shadowed in a blue rectangular in each figure. The mirror plane M$_a$ of the <100> direction is marked by solid line and the glide mirror plane $\widetilde{M}_b$ of the <010> direction is marked by dash line in figure a, the screw axis along the <001> direction is marked by symbol 2$_1$. **c,** Collection of the local photocurrent from a WTe$_2$ device in a Shockley-Ramo-type scheme. The solid grey lines are the weighting field lines of the device used in Figure 2a with the source contact set at 1 V and the drain contact set at 0 V. The red arrows represent the possible directions of local edge photocurrent. The simulations are performed using COMSOL. **d,** The distribution of Weyl nodes in bulk WTe$_2$ and the projections of Weyl nodes on (100), (010) and (110) surfaces. WTe$_2$ has eight Weyl nodes all locating in k$_c$ = 0 plane, each four symmetry-related nodes are referred as W$_1$ and W$_2$ respectively. The mirror symmetry related Weyl nodes with opposite chirality annihilate in pairs when projected onto (100) and (010) surfaces, but the Fermi arcs can survive in higher Miller-index surface BZs.

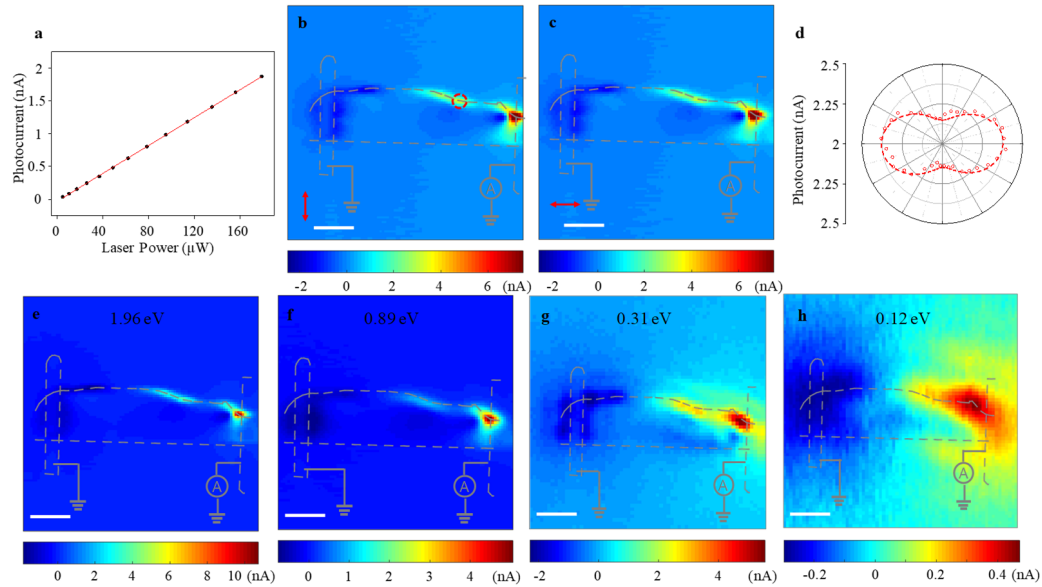

**Figure 4| Excitation power and polarization dependence of edge photocurrent responses of WTe$_2$. a**, Power dependence of the edge photocurrent responses with 1.96-eV pulse excitation. **b, c**, Scanning photocurrent microscopies under horizontal- (b) and vertical-polarized (c) light excitation respectively. The light polarization is marked by the red arrows. **d**, Polarization dependent photocurrent responses with light focused on the spot marked by a red circle in figure b. **e-h**, Scanning photocurrent microscopies with excitation photon energies of 1.96 eV (e), 0.89 eV (f), 0.31 eV (g) and 0.12 eV (h) respectively. All measurements were performed on the device shown in Figure 2a. The scale bars are all 8 μm.

**Acknowledgements**

Q. W. acknowledge Yimeng Wang from Renmin University of China and Dr. Yilin Lu for the help in the measurement of Raman spectroscopy. This work is supported by the National Natural Science Foundation of China (NSFC Grants Nos. 11704031, 11674013, 11734003, 11574029, 91750109), the National Key R&D Program of China (Grant No. 2016YFA0300600), the Strategic Priority Research Program of Chinese Academy of Sciences (Grant No. XDB30000000), K. S. is supported by the NSF Award No. NSF-EFMA-1741618. Q. W. is supported by Beijing Institute of Technology Research Fund Program for Young Scholars. D.S. is supported by the State Key Laboratory of Precision Measurement Technology and Instruments Fund for open topics.

**Author contributions**

D.S. and Q.W. conceived the experiment. J.Z. and X.L. fabricated the devices under the supervision of Q.W. and J.C., Y.H. and Q.W. carried out the photocurrent measurements. J.M. and J.L. contributed to the photocurrent measurements. Q.W., J.C. and Y.H. analyzed and simulated the data under supervision from Y.Y. and D.S., H.L. and D.Y. synthesized the bulk crystals under supervision from S.J. and Y.-G. S. K.S., J.C. and Y.Y. contributed to theoretical discussions. Q.W. and D.S. co-wrote the paper with input from all the authors.